\documentclass[aps,reprint,groupedaddress,showpacs,longbibliography,nofootinbib]{revtex4-1}
\usepackage{amssymb}
\usepackage[breaklinks,colorlinks,citecolor=green,urlcolor=blue,linkcolor=red]{hyperref}
\usepackage{graphicx}
\usepackage{epstopdf}
\usepackage{diagbox}
\usepackage{float}
\usepackage{subfigure}
\usepackage{multirow}
\usepackage{CJK}
\usepackage{tikz}
\usepackage{enumerate}

\AtBeginDocument{%
    \newwrite\bibnotes
    \def\bibnotesext{Notes.bib}
    \immediate\openout\bibnotes=\jobname\bibnotesext
    \immediate\write\bibnotes{@CONTROL{REVTEX41Control}}
    \immediate\write\bibnotes{@CONTROL{%
    apsrev41Control,author="08",editor="1",pages="1",title="0",year="1"}}
     \if@filesw
     \immediate\write\@auxout{\string\citation{apsrev41Control}}%
    \fi
}%

\begin{document}

\newcommand{\RN}[1]{\uppercase\expandafter{\romannumeral#1}}
\newcommand{\CPV}{$C\!P$V}
\newcommand{\CPA}{$C\!P$A}

% Use the \preprint command to place your local institutional report
% number in the upper righthand corner of the title page in preprint mode.
% Multiple \preprint commands are allowed.
% Use the 'preprintnumbers' class option to override journal defaults
% to display numbers if necessary
%\preprint{}

%Title of paper
\title{\boldmath Searching for $C\!P$ violation through two-dimensional angular distributions in four-body decays of bottom and charmed baryons}

% repeat the \author .. \affiliation  etc. as needed
% \email, \thanks, \homepage, \altaffiliation all apply to the current
% author. Explanatory text should go in the []'s, actual e-mail
% address or url should go in the {}'s for \email and \homepage.
% Please use the appropriate macro foreach each type of information

% \affiliation command applies to all authors since the last
% \affiliation command. The \affiliation command should follow the
% other information
% \affiliation can be followed by \email, \homepage, \thanks as well.
\author{Zhen-Hua Zhang}
%\email{zhangzh@usc.edu.cn}
%\homepage[]{Your web page}
%\thanks{}
\affiliation{School of Nuclear Science and Technology, University of South China, Hengyang, 421001, Hunan, China}

%Collaboration name if desired (requires use of superscriptaddress
%option in \documentclass). \noaffiliation is required (may also be
%used with the \author command).
%\collaboration can be followed by \email, \homepage, \thanks as well.
%\collaboration{}
%\noaffiliation

\date{\today}

\begin{abstract}
It is proposed to search for $C\!P$ violation through the two-fold angular distributions
in the four-body decays of bottom and charmed hadrons.
The two polar angles in the two-fold angular distributions are correlated,
to which the interferences of intermediate resonances are one important origin.
These interferences will leave tracks in the two-fold angular distributions, with which the $C\!P$ violation can be studied.
Special attention is paid to the case when all the intermediate resonances are different, which is unique to four-body decays.
It is suggested to look for $C\!P$ violation in  four-body decays such as $\Lambda_b^0\to p \pi^-\pi^+\pi^-$ through the analysis of the two-fold angular distributions.
The method proposed in this paper is also widely applicable to other four-body decays processes.
\end{abstract}

% insert suggested PACS numbers in braces on next line
%\pacs{}
% insert suggested keywords - APS authors don't need to do this
%\keywords{CP violation, heavy baryon,}

%\maketitle must follow title, authors, abstract, \pacs, and \keywords
\maketitle

% body of paper here - Use proper section commands
% References should be done using the \cite, \ref, and \label commands
%\section{}
% Put \label in argument of \section for cross-referencing
%\section{\label{}}
%\subsection{}
%\subsubsection{}

%%%%%%%%%%%%%%%%%%%%%%%%%%%%%%%%%%%%%%%%
%%%%%%%%%%%%%%%%%%%%%%%%%%%%%%%%%%%%%%%%
\section{introduction}
%%%%%%%%%%%%%%%%%%%%%%%%%%%%%%%%%%%%%%%%
%%%%%%%%%%%%%%%%%%%%%%%%%%%%%%%%%%%%%%%%
$C\!P$ violation (\CPV), as one of the cornerstones for the explanation of the matter-antimatter asymmetry of the Universe through baryogenesis \cite{Sakharov:1967dj},
is of great importance in the baryon sector both theoretically and experimentally.
Searches for \CPV ~have been carried out in the baryonic decay channels of $C\!P$-conjugate pairs, % of the baryons,
through direct measurements of the differences between the full partial decay widths\cite{CDF:2011ubb,CDF:2014pzb,LHCb:2018fly,LHCb:2017hwf},
the decay widths corresponding to part of the phase space \cite{LHCb:2019jyj,LHCb:2020zkk},
 the Triple-Product Asymmetries (TPAs) \cite{LHCb:2016yco,LHCb:2018fpt,LHCb:2019oke},
  and the decay asymmetry parameters \cite{FOCUS:2005vxq,BESIII:2022qax,Belle:2022uod},
or through other indirect techniques such as the amplitude analysis \cite{LHCb:2021enr}
and the energy test method \cite{LHCb:2019oke}, etc..
Puzzling enough, although it was first discovered almost sixty years ago in the neutral-kaon decays \cite{Christenson:1964fg},
\CPV ~has never been observed in the baryon sector,
despite that so many efforts have been paid.

Contrary to the baryonic cases, \CPV ~has been observed in the decays of $K$ and $D$ mesons \cite{Christenson:1964fg,LHCb:2019hro},
and in plenty of decay channels of $B$ mesons \cite{BaBar:2001ags,Belle:2001zzw,LHCb:2013syl,Workman:2022ynf},
some of which have quite large $C\!P$ asymmetries (\CPA s).
One such kind of examples are the large regional \CPA s observed in parts of the phase space for some three-body decays of B meson \cite{LHCb:2013ptu,LHCb:2014mir,LHCb:2019sus,LHCb:2022nyw},
in which the interfering effect between the nearby resonances (or the interference of resonance with the non-resonant part) is one important mechanism for the generating of large regional \CPA s \cite{Zhang:2013oqa,Bhattacharya:2013cvn,Cheng:2013dua,Cheng:2020ipp,Wei:2022zuf}.
Similar to  the $B$ meson cases, multi-body decays of bottom and charmed baryons are usually dominated by various intermediate resonances.
For example, $\Lambda_b^0\to p\pi^+\pi^-\pi^-$ is dominated by $N(1520)^0$ and $\rho(770)^0$ via ``branching'' decays $\Lambda_b^0\to N(1520)^0(\to p\pi^-)\rho(770)^0(\to \pi^+\pi^-)$ \cite{LHCb:2019jyj}.
The prevalent interferences between the amplitudes corresponding to various intermediate resonances
are expected to leave tracks in the angular distributions of the final particles, in which the \CPV ~is potentially hidden.

Based on the Cabibbo-Kobayashi-Maskawa mechanism of the Standard Model \cite{Cabibbo:1965zzb,Kobayashi:1973fv},
the \CPA s ~in some of the bottom and charmed baryon decays are expected to have similar magnitudes with those in the bottom and charmed meson decays,
respectively \cite{Lu:2009cm,Hsiao:2014mua,Shi:2019vus}.
However, the baryon decays usually suffer from substantial lower statistics comparing with the meson cases.
Hence one of the best strategies for the \CPV ~searching in the baryon sector is to start with the decay channels with the largest statistics.
The four-body decays such as $\Lambda_b^0\to p K^- \pi^+\pi^-$ and $\Lambda_b^0\to p\pi^+\pi^-\pi^-$ are one important type that fulfill this criteria.

Although \CPA s corresponding to angular distributions of final particles have been extensively investigated for the heavy baryon decays in the literature
\cite{Giri:2001ju,Bensalem:2002pz,Bensalem:2002ys,Bigi:2012ev,
Bigi:2017eni,Kang:2010td,Gronau:2015gha,Durieux:2016nqr,Shi:2019vus,Zhang:2021fdd,Zhang:2021sit,Geng:2022osc,Wang:2022tcm,Zhang:2022emj},
for the four-body decay cases, however, this area is largely unexplored.
As one kind of angular distribution correlated \CPA s, the TPA induced \CPA s have played an unique role in heavy meson and baryon decays \cite{Valencia:1988it,Dunietz:1990cj,Golowich:1988ig,Kayser:1989vw,Bensalem:2000hq,Durieux:2015zwa,Bensalem:2002pz,Bensalem:2002ys}.
In fact, the TPA induced \CPA s are especially suitable be studied in four-body decays of heavy baryons \cite{Gronau:2015gha,Shi:2019vus}.
Although it is in principle possible to study TPA and corresponding \CPA s in two- or three-body decays of heavy baryons \cite{Kayser:1989vw,Bensalem:2000hq,Bensalem:2002pz,Bensalem:2002ys},
this requires, however, that the mother baryons to be polarized, which is not possible in the current stage \cite{LHCb:2013hzx,CMS:2018wjk,LHCb:2020iux}.
In this paper, \CPV ~corresponding to angular distributions, which complements to that corresponding to TPA,
is investigated for the aforementioned four-body decays with the largest statistics.

%%%%%%%%%%%%%%%%%%%%%%%%%%%%%%%%%%%%%%%%
%%%%%%%%%%%%%%%%%%%%%%%%%%%%%%%%%%%%%%%%
\section{The two-fold angular distributions for four-body decays}
\begin{figure}[t]\center
\includegraphics[width=0.5\textwidth]{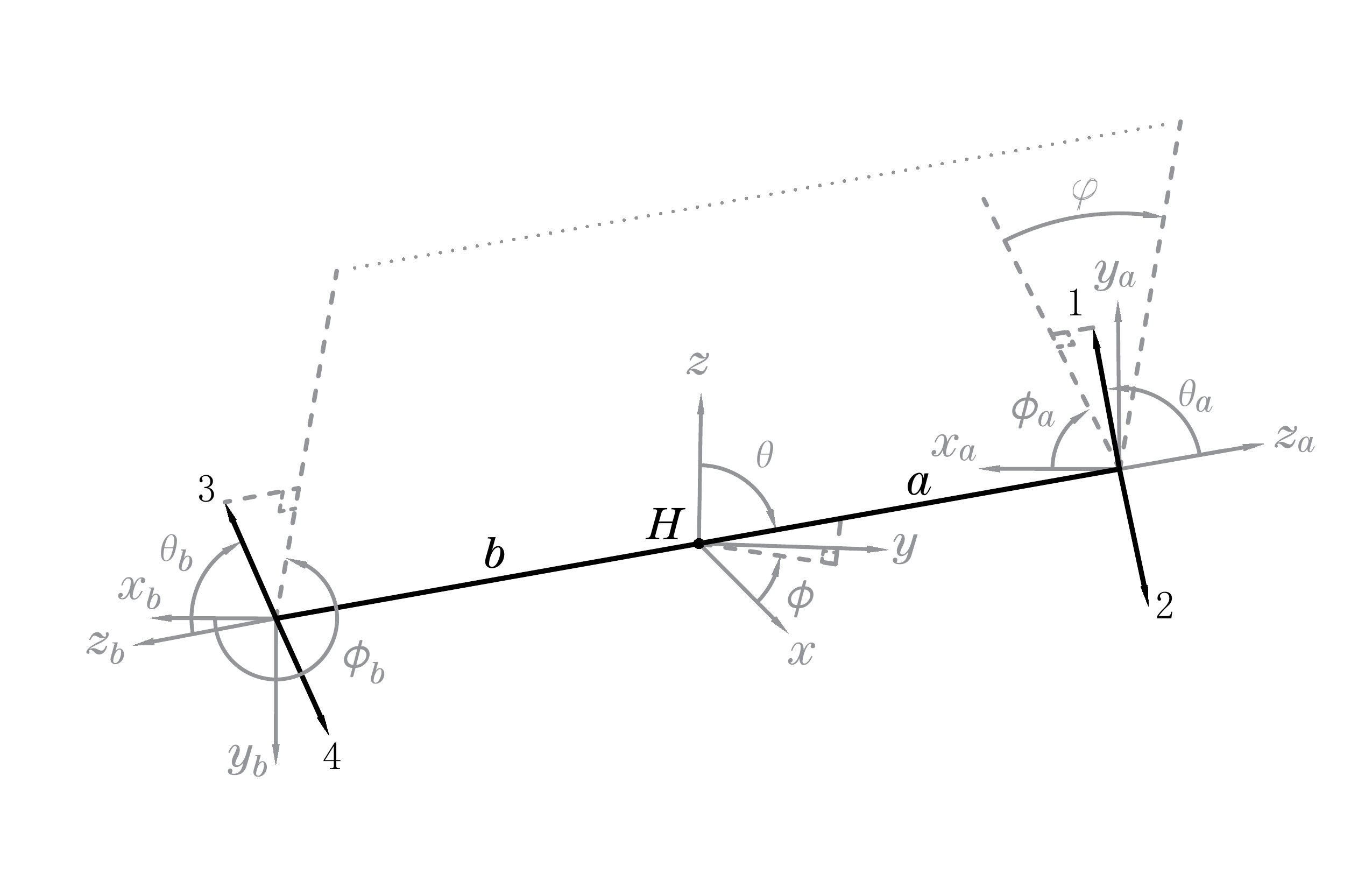}
\caption{Illustration of the kinematic variables for the four-body decay $H\to a (\to12) b (\to 34)$.
         The reference frames is defined according to the Jackson
         convention. Note that $\theta$ and $\phi$ are defined in the c. m. frame of $H$,
         while $\theta_{a(b)}$ and $\phi_{a(b)}$ are defined in the c. m. frame of $a(b)$.
         The angle $\varphi$ can be defined in any of the three frames.
         The invariant mass squared of 12 and 34, $s_{12}$ and $s_{34}$, which are Lorentz-invariant, are not shown in the figure.
         }
\label{fig:frames}
\end{figure}

%%%%%%%%%%%%%%%%%%%%%%%%%%%%%%%%%%%%%%%%
%%%%%%%%%%%%%%%%%%%%%%%%%%%%%%%%%%%%%%%%
To put on a more general ground, let us first consider a four-body decay process of the branching form $H\to a (\to12) b (\to 34)$,
where $H$ is the mother hadron, either baryon or meson, bottom or charmed, $a$ and $b$ are the intermediate resonances,  and 1234 label the four final particles.
The kinematical variables are illustrated in FIG. \ref{fig:frames}, where the Jackson convention is adopted for the reference frames \cite{Gottfried:1964nx}.
In the center-of mass (c. m.) frame of $H$, the $z$ axis is conveniently chosen as the normal to the production plane of $H$.
The polar and azimuthal angles of the momentum of $a$ in this frame is denoted as $\theta$ and $\phi$.
In the c. m. frame of $a$($b$), the $z_{a(b)}$ axis is chosen along the direction of the momentum of $H$.
The $x_a$ and $x_b$ ($y_a$ and $y_b$) axes will be chosen (anti-)aligned with each other.
The polar and the azimuthal angles of the particle 1(3) in the c. m. frame of $a$($b$) will be denoted as $\theta_{a(b)}$ and $\phi_{a(b)}$, respectively.
The angle from the decay plane of the $a\to 12$ to that of the $b\to34$, which is related to TPA, will be denoted as $\varphi$
\footnote{For the current situation, the TPA constructed from the momenta of the final particles can be defined as $A_{T}\equiv \frac{N(C_T>0)-N(C_T<0)}{N(C_T>0)+N(C_T<0)}$, with $C_T\equiv (\mathbf{p}_1\times\mathbf{p}_2)\cdot\mathbf{p}_3$, where $\mathbf{p}_j$ represent the momentum of particle $j$ in the c. m. of H.
Form FIG. \ref{fig:frames} one can see that the above definition is equivalent to $A_{T}=\frac{N(\sin\varphi>0)-N(\sin\varphi<0)}{N(\sin\varphi>0)+N(\sin\varphi<0)}$.
Consequently, the aforementioned TPA $A_T$ is sometimes also called as the up-down asymmetry is some references as the decay plane of the 12 system divide the space into the ``up'' part ($\sin\varphi>0$) and the ``down'' part ($\sin\varphi<0$) \cite{Delaunay:2013npa,Durieux:2015zwa}.}.
One can find from FIG. \ref{fig:frames} the relation $\varphi=2\pi-\phi_a-\phi_b$.

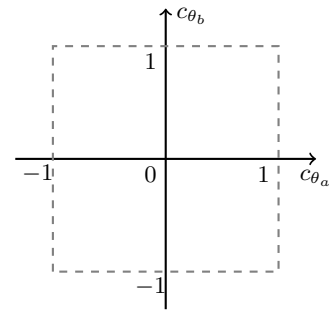
\begin{figure}[b]\centering
\begin{tikzpicture}
\draw[->, black, thick] (-2,0) -- (2,0);
\draw[->, black, thick] (0,-2) -- (0,2);
\draw[dashed, gray, thick] (-1.5,-1.5) rectangle(1.5,1.5);
 \node[] at(2,-0.25) {$c_{\theta_a}$};
 \node[] at(0.35,1.9) {$c_{\theta_b}$};
 \node[] at(-0.2,1.3) {1};
 \node[] at(-0.2,-0.2) {0};
 \node[] at(-0.2,-1.7) {$-1$};
 \node[] at(1.3,-0.2) {1};
 \node[] at(-1.7,-0.2) {$-1$};
\end{tikzpicture}
  \caption{Illustration of the 2DPS expanded by $c_{\theta_a}$ and $c_{\theta_b}$.}
    \label{FIG:4Region}
\end{figure}

The spin-averaged decay amplitude squared for unpolarized $H$,
{\it after integrating out the azimuthal angles $\phi_a$ and $\phi_b$,} will take the form
\footnote{The expression of the spin-averaged decay amplitude squared for polarized $H$ without integrating out the azimuthal angles are presented in Appendix \ref{sec:appA}.},
\begin{equation}\label{eq:AngDis}
  \int\overline{\left| \mathcal{A}\right|^2}d\varphi \propto \sum_{jl}\Gamma_{jl}(s_{12},s_{34})P_{j}(c_{\theta_a})P_{l}(c_{\theta_b})
\end{equation}
where $P_{j (l)}$ are the Legendre polynomials with $c_{\theta_{a(b)}}\equiv\cos\theta_{a(b)}$, and
\begin{equation}\label{eq:Gammajl}
  \Gamma_{jl}=
  \sum_{a,a',b,b'}
  \frac{\mathcal{W}^{(ab,a'b')}_{jl}\mathcal{G}^{(aa')}_j\mathcal{G}^{(bb')}_l}{\mathcal{I}_{a}\mathcal{I}_{a'}^\ast\mathcal{I}_{b}\mathcal{I}_{b'}^\ast},
\end{equation}
 $\mathcal{I}$'s are the reciprocals of the Breit-Wigner propagators, taking the form
 $\mathcal{I}_{a^{(\prime)}}=s_{12}-m_{a^{(\prime)}}^2+im_{a^{(\prime)}}\Gamma_{a^{(\prime)}}$ and
 $\mathcal{I}_{b^{(\prime)}}=s_{34}-m_{b^{(\prime)}}^2+im_{b^{(\prime)}}\Gamma_{b^{(\prime)}}$
 with $s_{12}$ and $s_{34}$ the invariant mass squared of the 12 and the 34 systems, and
\begin{eqnarray}
  \mathcal{W}^{(ab,a'b')}_{jl}&=&\sum_{\sigma\rho}(-)^{\sigma-s_a+\rho-s_b}\langle s_{a}-\sigma s_{a'}\sigma|j0\rangle \nonumber \\
  &&\times\langle s_{b}-\rho s_{b'}\rho|l0\rangle
  \mathcal{F}_{\sigma\rho}^{H\to ab}\mathcal{F}_{\sigma\rho}^{H\to a'b' \ast},
\end{eqnarray}
%\end{widetext}
\begin{equation}\label{eq:Gj}
  \mathcal{G}^{(aa')}_j=\sum_{\lambda_1\lambda_2}(-)^{s_a-\lambda_{12}}\langle s_{a}-\lambda_{12}s_{a'}\lambda_{12}|j0\rangle
  \mathcal{F}^{a\to 12}_{\lambda_1\lambda_2}\mathcal{F}^{a'\to 12\ast}_{\lambda_1\lambda_2},
\end{equation}
\begin{equation}\label{eq:Gl}
\mathcal{G}^{(bb')}_l=\sum_{\lambda_3\lambda_4}(-)^{s_b-\lambda_{34}}\langle s_{b}-\lambda_{34}s_{b'}\lambda_{34}|l0\rangle
  \mathcal{F}^{b\to34}_{\lambda_3\lambda_4}\mathcal{F}^{b'\to34\ast}_{\lambda_3\lambda_4},
\end{equation}
with all the $\mathcal{F}$'s being the decay amplitudes in the helicity form,
``$\langle\cdots |\cdots \rangle$'' being the Clebsh-Gordan coefficients,
$s_{a^{(\prime)}}$ and $s_{b^{(\prime)}}$ being the spins of $a^{(\prime)}$ and $b^{(\prime)}$,
$\sigma$ and $\rho$ being the helicities of $a^{(\prime)}$ and $b^{(\prime)}$ in the c. m. frame of $H$,
$\lambda_i$ ($i=1,\cdots,4$) being the helicity index of particle $i$ (defined in the c. m. frame of $a$ or $b$),
$\lambda_{ii'}=\lambda_i-\lambda_{i'}$.
The summations over $a$, $a'$, $b$, and $b'$ indicate that there may be more than one resonances either for $a$ or $b$.
Note that there are four kinematic variables left, which can be chosen as the invariant mass squared of $12$ and $34$, $s_{12}$ and $s_{34}$,
the polar angles $\theta_a$ and $\theta_b$.
When $s_{12}$ and $s_{34}$ are set fixed in certain ranges, we get a two-dimensional phase space (2DPS) expanded by $c_{\theta_a}$ and $c_{\theta_b}$,
which is illustrated in FIG. \ref{FIG:4Region}.

The following remarks to the two-fold angular distributions in Eq. (\ref{eq:AngDis}) are in order:

\begin{enumerate}[i)]
\item
  The two angles $\theta_a$ and $\theta_b$ are correlated, in the sense that
$\Gamma_{jl}$ can not be factorized into the form $\Gamma_{jl}=\xi_j\eta_l$.
If $\Gamma_{jl}$ were factorized, that would mean that one can simply integrate out either $\theta_a$ or $\theta_b$ without lost of any information,
so that the four-body decay will reduced effectively to three-body decays.
However, this is not true, because that the amplitudes are entangled in Eq. (\ref{eq:Gammajl}).
Among the origins for the entanglement of $j$ and $l$ in Eq. (\ref{eq:Gammajl}), the {\it genuine interferences} (GIs) between the amplitudes of $H\to a(\to12) b(\to34)$ and $H\to a'(\to12) b'(\to34)$,
where $a\neq a'$ {\it and} $b\neq b'$, are the most special ones.
The intermediate resonances of the two amplitudes are {\it all} different,
hence they are unique for the four-body decays and can not be accounted for by any two- or three-body decays.
In other word, GI is a two-fold interference, because it contains both the $a\leftrightarrow a'$ and $b\leftrightarrow b'$ interferences.
Anyway, it would be interesting to study the correlation between $\theta_a$ and $\theta_b$ both theoretically and experimentally.

\item
From the Clebsh-Gordan coefficients in Eq. (\ref{eq:Gj}) one can see that $j$ takes integer values from 0 to $2\max_a(s_a)$,
where $a$ runs over all the possible resonances.
Similarly, $l$ takes integer values from 0 to $2\max_b(s_b)$.
Note that the allowed values of $j$ and $l$ are determined by the {\it spin} of the resonances through the Clebsh-Gordan coefficients,
 not directly by the angular momentum between the final particles 1 and 2 (or 3 and 4).
 Though they equal with each other (the spin of the resonances and the angular momentum of 12 or 34) only when all the final particles are spin-0 ones.
\item
In principle, \CPV ~can show up in any of the $\Gamma_{jl}$'s through
the difference between $\Gamma_{jl}$ and its $C\!P$ conjugate  $\overline{\Gamma_{jl}}$.
However, it is more likely that \CPV ~shows in those $\Gamma_{jl}$'s which contain the resonant-interfering terms.
This is because that there may be, between the amplitudes corresponding to different resonances, non-perturbative strong phase differences,
which are crucial for the generating of large \CPA s.
Consequently, it is important to find the rules whether there are resonant-interfering terms in each $\Gamma_{jl}$ or not.
To this end, notice that the parity conservation, as well as the properties of the Clebsh-Gordan coefficients, imply from Eqs. (\ref{eq:Gj}) and (\ref{eq:Gl}) that  $\Pi_{a}\Pi_{a'}(-)^j=1$ and $\Pi_{b}\Pi_{b'}(-)^l=1$,
where $\Pi$'s represent the parities of the corresponding particles.
It follows that there are selection rules (SRs) for the presence of (non-)interference terms in $\Gamma_{jl}$, which fall into four situations:
(1) The non-interfering terms ($a=a'$, $b=b'$) for $H\to ab$ will present in $\Gamma_{jl}$ if $0\leq j \leq 2s_a$ and $0\leq l\leq2s_b$, and $j$ and $l$ are even.
(2) The singly interfering terms between $H \to ab$ and $H \to a'b$ will present in $\Gamma_{jl}$ if $|s_a-s_{a'}|\leq j\leq s_a+s_{a'}$,
 $\Pi_{a}\Pi_{a'}(-)^j$ is positive, $0\leq l\leq 2s_b$, and $l$ is even.
(3) The singly interfering terms between $H \to ab$ and $H \to ab'$ will present in $\Gamma_{jl}$ if $|s_b-s_{b'}|\leq l\leq s_b+s_{b'}$,
 $\Pi_{b}\Pi_{b'}(-)^l$ is positive, $0\leq j\leq s_a$, and $j$ is even.
(4) The GI terms will present in $\Gamma_{jl}$ if
$|s_a-s_{a'}|\leq j\leq s_a+s_{a'}$, and $|s_b-s_{b'}|\leq l\leq s_b+s_{b'}$, and $\Pi_{a}\Pi_{a'}(-)^j$ and $\Pi_b\Pi_{b'}(-)^l$ are positive.

\item
One special case is when the intermediate resonances have opposite parities in each of the two decay branches.
For example, suppose that there are two resonances at each decay branch of $a$ and $b$,
which are $R^{(+)}_a$, $R^{(-)}_a$ and $R^{(+)}_b$, $R^{(-)}_b$, respectively, with opposite parities indicated in the superscripts.
According to the SRs, the non-interfering terms, the singly interfering terms,
and the GI terms will show up in $\Gamma_{jl}$ when both $j$ and $l$ even, $j$ even $l$ odd or $j$ odd $l$ even, and both $j$ and $l$ odd, respectively.
Hence they are all well separated in $\Gamma_{jl}$'s.
\item
Each of the dynamical parameters $\Gamma_{jl}$'s in Eq. (\ref{eq:AngDis}) represents a degree of freedom for angular distributions,
which can be expressed as
\begin{equation}\label{eq:GammaExp}
  \Gamma_{jl}\propto\int \overline{|A|^2}P_j(c_{\theta_{a}})P_l(c_{\theta_{b}}) dc_{\theta_a}dc_{\theta_b}.
\end{equation}
Consequently, $\Gamma_{jl}$ can be determined experimentally according to
\begin{equation}\label{eq:GammaExp}
  \Gamma_{jl}\propto\sum_{i=1}^{N}P_j(c_{\theta_{ai}})P_l(c_{\theta_{bi}}),
\end{equation}
where $N$ is the total event yields, and $i$ labels each event.
The relative $jl$-th moment for the expansion of Eq. (\ref{eq:AngDis}) is then defined as
\begin{equation}\label{eq:Ajl}
  A^{jl}\equiv \frac{\Gamma_{jl}}{\Gamma_{00}}.
\end{equation}
 The \CPA s corresponding to $A^{jl}$ can then be obtained according to
\begin{equation}\label{eq:AjlCP}
  {A}_{CP}^{jl}\equiv\frac{1}{2}(A^{jl}-\overline{A^{jl}}),
\end{equation}
for $j\neq0$ and/or $l\neq0$, where $\overline{A^{jl}}$ is the same with ${A}_{CP}^{jl}$ but for the $C\!P$ conjugate process
\footnote{
\CPA s are alternatively defined as
$A_{CP}^{(jl)}\equiv \frac{\Gamma_{jl}-\overline{\Gamma_{jl}}}{\Gamma_{00}+\overline{\Gamma_{00}}}.$
For $j=0=l$, it reduces to the direct \CPA ~defined by the decay width:
$A_{CP}^{(00)}=A_{CP}^{\text{dir.}}$, as one can see that $\Gamma_{00}=N$,
and $\overline{\Gamma_{00}}=\overline{N}$.
}.

\item
Although very clean from the theoretical side, the disadvantage of the \CPA ~defined in Eq. (\ref{eq:AjlCP}) is also obvious:
the events located in different positions of the 2DPS are weighted differently, which complexes the experimental analysis of uncertainties.
To avoid this, we use an alternative definition for \CPA, which weights all the events  equally throughout the whole 2DPS (up to a signature).
To achieve this, one first notice that the contributions of each event to $\Gamma_{jl}$ are different in signature because of the factor $P_jP_l$ in Eq. (\ref{eq:GammaExp}).
Hence one can introduce a twisted $\Gamma_{jl}$ according to
\begin{equation}\label{eq:TwistGamma}
  \tilde{\Gamma}_{jl}\propto\sum_{i=1}^N \text{sgn} \left(P_J(\cos_{\theta_{ai}})P_l(\cos_{\theta_{bi}})\right).
\end{equation}
This is equivalent to dividing the 2DPS into $(j+1)\times(l+1)$ bins which are boarded by the zero lines of the Legendre polynomials $P_j$ and $P_l$,
and assign the event yields of each bin by a proper signature.
Eq. (\ref{eq:TwistGamma}) is then equivalent to
\begin{equation}\label{eq:TwistGamma2}
  \tilde{\Gamma}_{jl}\propto\sum_{i_{j}=0}^{j}\sum_{i_l=0}^{l}(-)^{i_j+i_l-l}N_{i_{j}i_{l}},
\end{equation}
where $N_{i_{j}i_{l}}$ is the event yields of the bin $i_{j}i_{l}$.
The twisted $jl$-th relative moment takes the form
\begin{equation}\label{eq:TwistAjl}
  \tilde{A}^{jl}=\frac{\tilde{\Gamma}_{jl}}{\Gamma_{00}}.
\end{equation}
The corresponding \CPA ~is then defined as
\footnote{Alternatively, \CPA s can be defined as
$
  \tilde{A}_{CP}^{(jl)}\equiv\frac{\tilde{\Gamma}_{jl}-\overline{\tilde{\Gamma}_{jl}}}{\Gamma_{00}+\overline{\Gamma_{00}}}.
$
}
\begin{equation}\label{eq:TwistAjlCP}
  \tilde{A}_{CP}^{jl}=\frac{1}{2}(\tilde{A}^{jl}-\overline{\tilde{A}^{jl}}).
\end{equation}

\item
Similar analysis is also applicable to the four-body sequential decays $H\to 1a$, $a\to 2b$, $b\to 34$.
It turns out that the SRs for the presence of different type of terms for the sequential decays are exactly the same with those for the branching ones.
A detailed analysis of this type of decays is presented in the appendix.
\end{enumerate}

%%%%%%%%%%%%%%%%%%%%%%%%%%%%%%%%%%%%%%%%
%%%%%%%%%%%%%%%%%%%%%%%%%%%%%%%%%%%%%%%%
\section{Applications}
%%%%%%%%%%%%%%%%%%%%%%%%%%%%%%%%%%%%%%%%
%%%%%%%%%%%%%%%%%%%%%%%%%%%%%%%%%%%%%%%%
As an application, we propose to look for \CPV ~in $\Lambda_b^0\to p \pi^+\pi^-\pi^-$ through the analysis of the two-fold anglular distributions in the 2DPS.
Previous experiments indicate that this decay channel is dominated by quasi-two-body decay $\Lambda_b^0\to N(1520)^0 \rho(770)^0$.
Hence there potentially are the $f_0(500)-\rho(770)^0$ and $N(1440)^0-N(1520)^0$ interferences
\footnote{The contribution of $f_0(500)$ is suppressed in the factorization approach.
Despite of this, similar analysis will apply as long as there are other contributions to the amplitudes interfering with $\rho(770)^0$, resonant or non-resonant.}.
Moreover, there can also be the GI terms such as the interference between $\Lambda_b^0\to N(1520)^0 f_0(500)$ and $\Lambda_b^0\to N(1440)^0 \rho(770)^0$.
The spin-parity of these two pairs of resonances are $0^+-1^-$ and $\left(\frac{1}{2}\right)^+-\left(\frac{3}{2}\right)^-$, respectively.
According to the SRs, both $j$ and $l$ can take values $0,1,2$.
The non-interfering terms, the singly interfering terms,
 and the GI terms are all perfectly separated in $\Gamma_{jl}$ because of the opposite parities of the two pairs of resonances.
The non-interfering terms will show up in $\Gamma_{jl}$ for $jl=00,02,20,22$;
the singly interfering terms will show up in $\Gamma_{jl}$ for $jl=01,10,12,21$;
while the GI terms will show up in and only in $\Gamma_{11}$.
This can be represented in a matrix form as
\begin{equation}\label{eq:illustration}
  (\Gamma_{jl}\!)\!\sim\!\!\left(\!\!
                \begin{array}{c|c|c}
                  \multirow{2}{*}{\text{Non-int}} & (N_{1440}N_{1520})|f|^2,   &  \multirow{2}{*}{\text{Non-int}}\\
                                 & (N_{1440}N_{1520})|\rho|^2 &                \\ \hline
                  (f\rho)|N_{1440}|^2, & \multirow{2}{*}{$(N_{1440}N_{1520}f\rho)_{GI}$} & \multirow{2}{*}{$(f\rho)|N_{1520}|^2$} \\
                  (f\rho)|N_{1520}|^2  &                              &                     \\ \hline
                  \text{Non-int} & (N_{1440}N_{1520})|\rho|^2 & \text{Non-int} \\
                \end{array}
              \!\!\right)\!.
\end{equation}
\CPV ~induced by the interference of the intermediate resonances can be ~embedded in any of the aforementioned five $\Gamma_{jl}$'s for $jl=01,10,12,21$, and 11,
which can be measured according to Eqs. (\ref{eq:Ajl}) and (\ref{eq:AjlCP}),
or alternatively, according to Eqs.(\ref{eq:TwistAjl}) and (\ref{eq:TwistAjlCP}).

\begin{figure}[t]\centering
\begin{tikzpicture}
\draw[scale=0.5,shift ={(-3 ,3)}, very thin] (-1,-1) rectangle (1,1);
\filldraw[scale=0.5,shift ={(0 ,3)},draw=gray!40,fill=gray!40] (-1,-1) rectangle (0,1);
\draw[scale=0.5,shift ={(0 ,3)},thick] (-1,-1) rectangle (1,1); \draw [scale=0.5,dashed,shift ={(0,3)}] (0,-1) -- (0,1);
\filldraw[scale=0.5,shift ={(3 ,3)},draw=gray!40,fill=gray!40] (-0.58,-1) rectangle (0.58,1);
\draw[scale=0.5,shift ={(3 ,3)}, very thin] (-1,-1) rectangle (1,1); \draw [scale=0.5,dashed,shift ={(3,3)}] (-0.58,-1) -- (-0.58,1) (0.58,-1) -- (0.58,1);

\filldraw[scale=0.5,shift ={(-3 ,0)},draw=gray!40,fill=gray!40] (-1,-1) rectangle (1,0);
\draw[scale=0.5,shift ={(-3 ,0)},thick] (-1,-1) rectangle (1,1);\draw [scale=0.5,dashed,shift ={(-3,0)}] (-1,0) -- (1,0);

\filldraw[scale=0.5,draw=gray!40,fill=gray!40] (-1,0) rectangle (0,1) (0,-1) rectangle (1,0);
\draw[scale=0.5, very thick] (-1,-1) rectangle (1,1);\draw [scale=0.5,dashed] (-1,0) -- (1,0)  (0,-1) -- (0,1);

\filldraw[scale=0.5,shift ={(3 ,0)},draw=gray!40,fill=gray!40] (-1,-1) rectangle (-0.58,0) (-0.58,0) rectangle (0.58,1) (0.58,-1) rectangle (1,0);
\draw[scale=0.5,shift ={(3 ,0)},thick] (-1,-1) rectangle (1,1);\draw [scale=0.5,dashed,shift ={(3,0)}] (-0.58,-1) -- (-0.58,1) (0.58,-1) -- (0.58,1) (-1,0) -- (1,0);
\filldraw[scale=0.5,shift ={(-3 ,-3)},draw=gray!40,fill=gray!40] (-1,-0.58) rectangle (1,0.58);
\draw[scale=0.5,shift ={(-3 ,-3)}, very thin] (-1,-1) rectangle (1,1);\draw [scale=0.5,dashed,shift ={(-3,-3)}] (-1,-0.58) -- (1,-0.58) (-1,0.58) -- (1,0.58);
\filldraw[scale=0.5,shift ={(0,-3)},draw=gray!40,fill=gray!40] (-1,-1) rectangle (0,-0.58) (0,-0.58) rectangle (1,0.58) (-1,0.58) rectangle (0,1);
\draw[scale=0.5,shift ={(0 ,-3)},thick] (-1,-1) rectangle (1,1); \draw [scale=0.5,dashed,shift ={(0,-3)}] (-1,-0.58) -- (1,-0.58) (-1,0.58) -- (1,0.58) (0,-1) -- (0,1);
\filldraw[scale=0.5,shift ={(3,-3)},draw=gray!40,fill=gray!40] (-1,-0.58) rectangle (-0.58,0.58) (-0.58,0.58) rectangle (0.58,1) (-0.58,-1) rectangle (0.58,-0.58) (0.58,-0.58) rectangle (1,0.58);
\draw[scale=0.5,shift ={(3 ,-3)}, very thin] (-1,-1) rectangle (1,1); \draw [scale=0.5,dashed,shift ={(3,-3)}] (-1,-0.58) -- (1,-0.58) (-1,0.58) -- (1,0.58) (-0.58,-1) -- (-0.58,1) (0.58,-1) -- (0.58,1);
\end{tikzpicture}
  \caption{Bin divisions of the 2DPS for the measurements of $\tilde{\Gamma}^{jl}$'s, $\tilde{A}^{jl}$'s, and $\tilde{A}^{jl}_{CP}$'s in $\Lambda_b^0\to p \pi^+\pi^-\pi^-$.
  There are nine ways of bin-dividing in total, each of which corresponds to the measurements of
  the aforementioned observables for $j$ and $l$ equals to the row and the columns in this figure.
  The bin divisions corresponding to the singly interfering terms are emphasized with thick squares (for $jl=01$, $10$, $12$, $21$),
  while the one corresponding to the GI terms is emphasized with even thicker one in the center of this figure (for $jl=11$).
  The padding in each bin indicates the signature in front of the corresponding event yields in Eq.
  (\ref{eq:TwistGamma2}) (white for positive and gray for negative).}
    \label{FIG:BinDiv}
\end{figure}
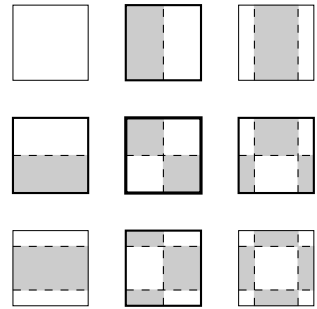

The bin divisions for the measurements of all the nine $\tilde{A}^{jl}$'s are illustrated in FIG. \ref{FIG:BinDiv}.
Since only $\Gamma_{11}$ contains the GI terms for the four-body decays, it deserves special attentions.
To measure it, one first needs to divide the 2DPS into four bins, which is illustrated as the one in the center of FIG. \ref{FIG:BinDiv}.
The $11$-th relative moment $\tilde{A}^{11}$, which can also be called as the two-fold forward-backward asymmetry (TFFBA), is then measured according to
\begin{equation}%\label{}
  \tilde{A}^{11}=\frac{\left(N_{I}-N_{I\!\!I}+N_{I\!\!I\!\!I}-N_{I\!V}\right)}{N},
\end{equation}
where the subscripts in the event yields denote the four quadrants according to the bin division shown in the center of FIG. \ref{FIG:BinDiv}.
One immediately obtains the \CPA ~$\tilde{A}^{11}_{CP}$ according to Eq. (\ref{eq:TwistAjlCP}) with $\tilde{A}^{11}$ and $\overline{\tilde{A}^{11}}$ at hand.

Similar 2DPS analysis can be performed in sequential decays.
For example, the four-body decay $\Lambda_b^0\to p \pi^+\pi^-\pi^-$ through the decay chains
$\Lambda_b^0\to p a_1(1260)^-$, $a_1(1260)^-\to\rho(770)^0\pi^-$, $\rho(770)^0\to \pi^+\pi^-$,
and $\Lambda_b^0\to N^{\ast+} \pi^-$, $N^{\ast+}\to\Delta^{++}\pi^-$, $\Delta^{++}\to p\pi^+$, which are also dominant, are suitable for the 2DPS analysis.
Take the former one as an example, potential interferences include those of
 $a_1(1260)^- - \pi(1300)^- $ and $\rho(770)^0-f(500)^0$, the spin-parities of which are respectively $1^+ - 0^-$ and $1^- -0^+$.
Hence both of $j$ and $l$ take values 0, 1, and 2.
Again, the singly interfering terms will show up in $jl=01$, 10, 02, 20, while the GI terms will show up in $jl=11$.
Note that there are no apparent suppressions for both $\pi(1300)^-$ and $f(500)^0$.
In this sense, this sequential decay
is even more suitable for the \CPV ~searching through the analysis of the two-fold angular distributions,
though there are potentially contaminations from other resonances such as $a_2(1320)^-$.

With a marginal generalization, the method proposed in this paper is also applicable to other cases.
One such example is the decay channel $\Lambda_c^+\to \Lambda^0 \overline{K^0}\pi^+$,
where the $\overline{K^0}\pi^+$ can decay from $K^*(892)^+$, and $\Lambda^0$ is reconstructed according to $\Lambda^0\to p \pi^-$.
Since $\Lambda^0\to p \pi$ is a weak process, SRs are no longer applicable.
However, one can equivalently view the parity-conserving and  -violating parts of the amplitudes
 of $\Lambda^0\to p \pi$ as amplitudes of two spin-half particles with opposite parities decaying into $p \pi$.
The interference of the parity-conserving and  -violating parts of the amplitudes,
together with that of $K_0^\ast(700)-K^\ast(892)$,
make up three angular distribution and corresponding \CPA ~observables (corresponding to $jl=01,10,11$) defined according to Eqs. (\ref{eq:TwistAjl}) or (\ref{eq:TwistAjlCP}),
where $\tilde{A}^{10}$ is in fact the decay asymmetry parameter, $\tilde{A}^{01}$ represents the forward-backward asymmetry in the $K^*(892)$ branch,
while $\tilde{A}_{11}$ contains the GI terms.
It would be interesting to measure the TFFBA $\tilde{A}_{11}$ and the corresponding \CPA ~$\tilde{A}^{11}_{CP}$.

%%%%%%%%%%%%%%%%%%%%%%%%%%%%%%%%%%%%%%%%
%%%%%%%%%%%%%%%%%%%%%%%%%%%%%%%%%%%%%%%%
\section{Summary}
%%%%%%%%%%%%%%%%%%%%%%%%%%%%%%%%%%%%%%%%
%%%%%%%%%%%%%%%%%%%%%%%%%%%%%%%%%%%%%%%%
\CPV ~corresponding to the two-fold angular distributions in four-body decays is analyzed.
The interferences of intermediate resonances may generate large \CPA s corresponding to the two-fold angular distributions,
 which have never been studied experimentally.
We propose to search for \CPV ~through the analysis of the two-fold angular distributions in the decay channels such as $\Lambda_b^0\to p \pi^+\pi^-\pi^-$ and $\Lambda_c^+\to\Lambda^0(\to p \pi)\overline{K^0} \pi^+ $.
This method is quite general, which can be used in a wide class of four-body decays of bottom and charmed hadrons, and are not limited to \CPV ~studies.

\begin{acknowledgments}
I thank Wenbin Qian for the constant inspiring discussions.
I also thank Long-Ke Li  and Jia-Jia Qin for helpful discussions and valuable suggestions on the manuscript.
Moreover, I would like to express my appreciation to the Referee for the constructive comments and suggestions, which help me a lot on the improvement of the manuscript.
This work was supported by National Natural Science Foundation of China under Grants No. 12192261, and Natural Science Foundation of
Hunan Province under Grants No. 2022JJ30483.
\end{acknowledgments}

\appendix
%%%%%%%%%%%%%%%%%%%%%%%%%%%%%%%%%%%%%%%%
%%%%%%%%%%%%%%%%%%%%%%%%%%%%%%%%%%%%%%%%
\section{\boldmath decay amplitude squared without integrating out $\varphi$ \label{sec:appA}}
%%%%%%%%%%%%%%%%%%%%%%%%%%%%%%%%%%%%%%%%
%%%%%%%%%%%%%%%%%%%%%%%%%%%%%%%%%%%%%%%%
We present here the decay amplitude squared without integrating out $\varphi$, and without the assumption of unpolarized $H$, which reads
\begin{equation}\label{eq:AngDisPor}
  \overline{\left| \mathcal{A}\right|^2} \propto \sum\gamma^{jl}_{\sigma_{a'}\sigma_{a}\sigma_{b'}\sigma_{b}}d^{j}_{\sigma_{a'a},0}({\theta_a})d^{l}_{\sigma_{b'b},0}({\theta_b}) e^{i\sigma_{aa'}\phi_a}e^{i\sigma_{bb'}\phi_b},
\end{equation}
where $\sum\equiv\sum_{aa'bb'}\sum_{\sigma_a\sigma_{a'}\sigma_b\sigma_{b'}}\sum_{jl}$, $\sigma_{a'a}=\sigma_{a'}-\sigma_a$, $\sigma_{b'b}=\sigma_{b'}-\sigma_b$, and
\begin{equation}%
  \gamma^{jl}_{\sigma_{a'}\sigma_{a}\sigma_{b'}\sigma_{b}}=
  \frac{{w}^{(ab,a'b')jl}_{\sigma_{a'}\sigma_{a}\sigma_{b'}\sigma_{b}}\mathcal{G}^{(aa')}_j\mathcal{G}^{(bb')}_l}{\mathcal{I}_{a}\mathcal{I}_{a'}^\ast\mathcal{I}_{b}\mathcal{I}_{b'}^\ast},
\end{equation}
 with
\begin{eqnarray}
  {w}^{(ab,a'b')jl}_{\sigma_{a'}\sigma_{a}\sigma_{b'}\sigma_{b}}&&=\langle s_{a}-\sigma_a s_{a'}\sigma_{a'}|j\sigma_{a'a}\rangle \langle s_{b}-\sigma_b s_{b'}\sigma_{b'}|l\sigma_{b'b}\rangle\nonumber \\
  &&\times (-)^{\sigma_{a}-s_a+\sigma_{b}-s_b}
  \mathcal{F}^{H\to ab}_{\sigma_{a}\sigma_{b}}\mathcal{F}^{H\to a'b' \ast}_{\sigma_{a'}\sigma_{b'}} P_{\sigma_{ab},\sigma_{a'b'}}(\theta),\nonumber\\
\end{eqnarray}
where $P_{\sigma_{ab},\sigma_{a'b'}}(\theta)$ describes the polarization of $H$ with $\sigma_{a^{(\prime)}b^{(\prime)}}=\sigma_{a^{(\prime)}}-\sigma_{b^{(\prime)}}$.
Note that we have set $\phi=0$ in Eq. (\ref{eq:AngDisPor}).
One can also set either $\varphi_{a}=0$ or $\varphi_{b}=0$ freely, which we did not do here.

Concerning the polarization of $H$, the only relevant $H$ in practice are the spin-half baryons such as $\Lambda_b$.
It is well known that $H$ can only polarize along the normal to the production plane due to the constraint of the parity symmetry in the producing process.
The factor $P_{\sigma_{ab},\sigma_{a'b'}}(\theta)$ then takes the form
\begin{equation}%\label{}
  P(\theta)=\frac{1}{2}\left(1+\left(\begin{array}{c c} \cos\theta & -\sin\theta\\ -\sin\theta & -\cos\theta
  \end{array}\right)P_z\right),
\end{equation}
with $P_z$ the polarization of $H$ along the $z$ axis.
For unpolarized $H$, $P_z=0$, as observed for the $\Lambda_b$ case in $pp$ collision by LHCb and CMS at a few percents level \cite{LHCb:2013hzx,CMS:2018wjk,LHCb:2020iux}, $P_{\sigma_{ab},\sigma_{a'b'}}(\theta)$ reduce to $\delta_{\sigma_{ab},\sigma_{a'b'}}/2$, so that $\sigma_{aa'}=\sigma_{bb'}$.
The last two factors in Eq. (\ref{eq:AngDisPor}) then reduce to $e^{-i\sigma_{aa'}\varphi}$.

A simultaneous analysis of the correlation of $\theta_a$, $\theta_b$, and $\varphi$ may give us deeper insights in $C\!P$V.
Take again the decay $\Lambda_b^0\to p \pi^+\pi^-\pi^-$ as an example, according to the SRs, %beside the one proportional to $\cos\theta_a\cos\theta_b$ in the main text,
the presence of the GI will also generate terms (with $j=1=l$) which are proportional to $d^1_{1,0}(\theta_a)d^1_{1,0}(\theta_b)\cos\varphi\sim\sin\theta_a\sin\theta_b\sin\varphi$ and $d^1_{1,0}(\theta_a)d^1_{1,0}(\theta_b)\sin\varphi\sim\sin\theta_a\sin\theta_b\cos\varphi$, respectively.
The simultaneous analysis can be performed by fitting the aforementioned two factors.
%Though applicable in principle, it should be point out, however, that the relative low statistics may result in large uncertainties to the final results of the simultaneous analysis.

Besides a simultaneous analysis, an analysis of solely the $\varphi$ dependence is simpler.
For the two aforementioned GI terms, the former has to do with TPA, and the related $C\!P$V can be studied accordingly.
While the latter can be described by the Left-Right Asymmetry (LRA):
\begin{equation}%\label{}
  A^{LR}=\frac{N_L-N_R}{N_L+N_R},
\end{equation}
where $N_{R/L}$ are the event yields defined by $N_{L/R}\equiv N(\cos\phi \gtrless 0)$.
The corresponding $C\!P$V can be studied through the LRA induced $C\!P$ asymmetry
\begin{equation}%\label{}
  A_{CP}^{LR}=\frac{1}{2}(A^{LR}-\overline{A}^{LR}),
\end{equation}
 %or alternatively, through \begin{equation} A_{CP}^{(LR)}=\frac{A^{LR}+\overline{A}^{LR}}{A^{LR}-\overline{A}^{LR}},\end{equation}
where $\overline{A}^{LR}$ is the LRA of the $C\!P$ conjugate process.

%%%%%%%%%%%%%%%%%%%%%%%%%%%%%%%%%%%%%%%%
%%%%%%%%%%%%%%%%%%%%%%%%%%%%%%%%%%%%%%%%
\section{\boldmath Two-dimensional angular distributions for $H\to 1a(\to 2b(\to 34))$}
%%%%%%%%%%%%%%%%%%%%%%%%%%%%%%%%%%%%%%%%
%%%%%%%%%%%%%%%%%%%%%%%%%%%%%%%%%%%%%%%%
For completeness, we present here a brief discussion on four-body sequential decays of the form $H\to 1a$, $a\to 2b$, $b\to 34$.
The spin-averaged decay amplitude squared can be expressed as
\begin{equation}%\label{}
  \int\overline{|\mathcal{A}|^2} d\varphi\propto\sum_{jl}\hat{\Gamma}_{jl}(s_{234},s_{34})P_j(c_{\hat{\theta}_a})P_{l}(c_{\hat{\theta}_b}),
\end{equation}
where $s_{234}$ are the invariant mass squared of the 234 system,
the angle $\hat{\theta}_{a(b)}$ is the polar angle of the momentum of $2 (3)$ in the c. m. frame of $a (b)$,
where $z_a$ and $z_b$ are defined respectively in the c. m. frame of $a$ and $b$ in a similar manner with those in the main text, and
\begin{equation}%\label{}
  \hat{\Gamma}_{jl}=
  \sum_{a,a'}\sum_{b,b'}
  \frac{\hat{\mathcal{W}}^{(a,a')}_{j}\hat{\mathcal{G}}^{(ab,a'b')}_{jl}\mathcal{G}^{(bb')}_l}{\mathcal{I}_{a}\mathcal{I}_{a'}^\ast\mathcal{I}_{b}\mathcal{I}_{b'}^\ast},
\end{equation}
where $\mathcal{I}_{a^{(\prime)}}=s_{234}-m_{a^{(\prime)}}^2+im_{a^{(\prime)}}\Gamma_{a^{(\prime)}}$ for now, and
\begin{equation}
  \hat{\mathcal{W}}^{(a,a')}_{j}=\sum_{\lambda_1\sigma}(-)^{\sigma_a-s_a}\langle s_{a}-\sigma s_{a'}\sigma|j0\rangle
  \mathcal{F}_{\lambda_1\sigma}^{H\to 1a}\mathcal{F}_{\lambda_1\sigma}^{H\to 1a' \ast},
\end{equation}
\begin{eqnarray}%\label{eq:Gj}
  \hat{\mathcal{G}}^{(ab,a'b')}_{jl}&=&\sum_{\lambda_2\rho}(-)^{s_a+s_b-\lambda_{2}}
  \langle s_{a}(\rho-\lambda_{2})s_{a'}(\lambda_{2}-\rho)|j0\rangle\nonumber\\&& \times\langle s_{b}-\rho s_{b'}\rho|l0\rangle
  \mathcal{F}^{a\to 2b}_{\lambda_2\rho}\mathcal{F}^{a'\to 2b'\ast}_{\lambda_2\rho}.
\end{eqnarray}
Note that the azimuthal angles have also been integrated out.

Suppose that $a\to 2b$ and $b\to34$ are strong processes,
there comes the non-zeroness conditions for $\hat{\mathcal{G}}^{(ab,a'b')}_{jl}$ and $\mathcal{G}^{(bb')}_l$ followed by the parity symmetry and the properties of the Clebsh-Gordan coefficients,
which can be organized as
i) $0\leq j\leq 2\max_{a}(s_{a})$, $0\leq l\leq 2\max_{b}(s_{b})$
ii) $\Pi_b\Pi_{b'}(-)^l$ is positive,
iii) $\Pi_a\Pi_{a'}(-)^j\Pi_b\Pi_{b'}(-)^l$ is positive.
One see that the conditions are in fact the same with the branching decay $H\to a(\to 12) b(\to 34)$.
Consequently, SRs for the presence of different kinds of terms in $\hat{\Gamma}_{jl}$ will also be exactly the same.
The non-interfering terms ($a=a'$ and $b=b'$) will show up in $\hat{\Gamma}_{jl}$ if $0\leq j \leq 2s_a$ and $0\leq l\leq2s_b$, and $j$ and $l$ even.
The $a\leftrightarrow a'$ interfering term ($a\neq a'$ and $b=b'$) will show up in $\hat{\Gamma}_{jl}$ if $|s_a-s_{a'}|\leq j\leq s_a+s_{a'}$, $\Pi_{a}\Pi_{a'}(-)^j$ positive, $0\leq l\leq 2s_b$, and $l$ even.
The $b\leftrightarrow b'$ interfering term ($b\neq b'$ and $a=a'$) will show up in $\hat{\Gamma}_{jl}$ if $\Gamma_{jl}$ if $|s_b-s_{b'}|\leq l\leq s_b+s_{b'}$, $\Pi_{b}\Pi_{b'}(-)^l$ positive, $0\leq j\leq 2s_a$, and $j$ even.
The $a\leftrightarrow a'$ and $b\leftrightarrow b'$ interfering term ($a\neq a'$ and $b\neq b'$) will show up in $\hat{\Gamma}_{jl}$ if $|s_a-s_{a'}|\leq j\leq s_a+s_{a'}$, $|s_b-s_{b'}|\leq l\leq s_b+s_{b'}$, and both $\Pi_a\Pi_{a'}(-)^j$ and $\Pi_b\Pi_{b'}(-)^l$ positive.
Especially, if the parities of $a$ and $a'$ as well as $b$ and $b'$ are opposite, all the aforementioned terms will be again well separated in $\hat{\Gamma}_{jl}$.

\bibliography{zzhbib}

\end{document}